\documentclass[preprint,11pt]{elsarticle}

\usepackage{hyperref}
\usepackage{braket}
\usepackage{amssymb}
\usepackage{amsmath}
\usepackage{mathpazo}
\usepackage{fullpage}

\journal{XXXXXXXXXX}

\begin{document}
\begin{frontmatter}


\title{Black hole thermodynamics from decoherence}


\author[label1,label2]{Xiao-Kan Guo}
\ead{kankuohsiao@whu.edu.cn}
\address[label1]{University of Chinese Academy of Sciences, Beijing 100049, China}
\address[label2]{Wuhan Institute of Physics and Mathematics, Chinese Academy of Sciences, Wuhan 430071, China}

\begin{abstract}
We present an approach to the four laws of black hole thermodynamics by utilizing the thermodynamics of quantum coherence. Firstly, Hawking effect is attributed to the decoherence of the two-mode squeezed state in a black hole spacetime. Then use is made of the relative entropy between undecohered and decohered squeezed states whose monotonicity gives the zeroth and the second law, while the first law can be obtained either by the vanishing of the first derivative of relative entropy or by studying the effective thermal model generated by the modular Hamiltonian. Futhermore, information-theoretic arguments give a Planck's form of the third law of black hole thermodynamics. With this approach we can understand the laboratory analogues of black holes solely by quantum theory, and find a way to detect the  thermodynamics of black holes produced in colliders.

\end{abstract}


\end{frontmatter}
\section{Introduction}
\label{1}
After more than forty years of development, black hole thermodynamics remains to be an impetus to the current researches in black hole physics. The classical four laws of black hole mechanics \cite{BF01645742} are derived from the classical geometries of the event horizons, and are just similar to the four laws of thermodynamics. It is the discovery of Hawking radiation \cite{BF02345020} that turns the black hole mechanics into black hole thermodynamics and fixes the Bekenstein-Hawking formula for the black hole entropy \cite{PhysRevD.13.191}.

Hawking radiation is purely thermal and leads to the black hole information problem that initial pure quantum states can evaporates into mixed states \cite{PhysRevD.14.2460}, in other words, the \textquotedblleft loss of quantum coherence" \cite{Wald94}. Then what exactly does \textquotedblleft quantum coherence" mean? In an insightful paper \cite{PhysRevD.42.3413}, Grishchuk and Sidorov pointed out that the particle creation process described by the Bogoliubov transformation between in-out modes, as in \cite{BF02345020}, is equivalent to acting on the initial vacuum state a two-mode squeeze operator. The resulting out state is a two-mode squeezed state that precisely describes the \textquotedblleft quantum coherence". Kiefer \cite{0264-9381-18-22-101} then studied the behavier of the explicit squeezed state in the Schwarzschild spacetime and found that the thermal nature of Hawking effect is due to the observational coarse-graining of the quickly rotating squeezing angle which in effect makes the off-diagonal components of the reduced density matrix vanish, that is, \emph{decoherence}. From this perspective, the entropy generated by coarse-graining or decoherence gives the exact formula of Hawking temperature, while the whole state remains pure and hence there is no information loss.

Now that the thermal nature of Hawking effect can be attributed to decoherence, and given the importance of Hawking effect in black hole thermodynamics, one expects to find relations between the thermodynamics of quantum coherence \cite{1308.1245} and the black hole thermodynamics. In this paper, the four laws of black hole thermodynamics are derived from the coherent nature of Hawking radiation. It is argued that the four laws of black hole thermodynamics, as in the case of Hawking effect, can be treated as solely quantum effects of the decoherence. The advantage of doing this is that we can now understand the laboratory analogues of black hole phenomena \cite{0953-4075-45-16-163001} effectively in terms of quantum physics only, without referring to many concepts of gravitation, and hence the word {\it analogue} can really be replaced by {\it simulation}, which is inappropriate in the conventional sense \cite{zcz}. 

In next section a brief review of Hawking radiation from decoherence is given at first and then we argue that the area law of black hole entropy can be derived from decoherence. In section \ref{3}, the first three laws of black hole thermodynamics are derived from the propeties of the relative entropy that measures quantum coherence \cite{PhysRevLett.113.140401} and an information-theoretic speculation of the third law is given. In section \ref{4}, some discussions are given. In the appendix more discussions on the data compression for black holes are presented.
\section{The setup}\label{2}
Consider for simplicity a massless scalar field in Schwarzschild spacetime. The infalling vaccum state $\ket{0_-}$ evolves into the outcoming vacuum state $\ket{0_+}$ which can be related to $\ket{0_-}$ via Bogoliubov transformation \cite{BF02345020}. This is equivalent to acting on $\ket{0_-}$ a two-mode squeeze operator for each mode $k$ \cite{PhysRevD.42.3413,0264-9381-10-11-012},
\begin{equation}\label{f1}
\ket{0_+}=S(r_k,\phi_k)\ket{0_-}\equiv\ket{ss_k},
\end{equation}
where 
\begin{equation}\label{iso}
S(r_k,\phi_k)=\exp\{r_k(a_k a_{-k}e^{-2i\phi_k}-a^{\dagger}_k a^{\dagger}_{-k}e^{2i\phi_k})\}
\end{equation}
is the two-mode squeeze operator with squeeze factor $r_k$ and squeeze angle $\phi_k$. The annihilation/creation operators are related to the antiparticle/particle pair creation process. For the total state including all modes, one can use the composite two-mode squeeze operator $S(r,\phi)=\prod_kS(r_k,\phi_k)$ to get
\begin{equation}\label{f33}
\ket{\psi,t}=S(r,\phi)\ket{0_-}=\prod_k\otimes\ket{ss_k}.
\end{equation}
The density matrix $\rho=\ket{\psi,t}\bra{\psi,t}$ can be readily calculated. After that the density matrix $\rho$ decoheres into the diagonal reduced density matrix $\rho_{red}$ for the reason that the timescale of the exchange of squeezing direction is much shorter than the observational time due to the quick rotation of squeeze angles such that the off-diagonal components vanish \cite{0264-9381-18-22-101}. The von Neumann entropy with respect to $\rho_{red}$ is then
\begin{align}
S=&-\text{tr}(\rho_{red}\ln\rho_{red})=\nonumber\\
=&\sum_k[(1+\sinh^2 r_k)\ln(1+\sinh^2 r_k)-\sinh^2 r_k\ln\sinh^2 r_k].\label{f44}
\end{align}
When $r_k\gg1$ such that the coarse-graining is valid, with the understanding that the relative radiation rate of mode $k$ is
\begin{equation}\label{555}
\Gamma_k=\tanh^2r_k=\exp[-8\pi\omega M],
\end{equation}
we have then for Hawking quanta in a volume $V$,
\begin{equation}\label{f4}
S\rightarrow \frac{2\pi^2}{45}T^3_{\text{BH}}V,
\end{equation}
which is exactly the entropy of Hawking radiation from Schwarzschild black hole with $T_{\text{BH}}=(8\pi M)^{-1}$ \cite{0264-9381-18-22-101,PhysRevD.13.191}.

To proceed the discussions on black hole thermodynamics, it is important to find the entropy of a black hole, to wit, Bekenstein-Hawking entropy plus possible logarithmic corrections. A microscopic description of this issue is desirable, however, since the squeezed state \eqref{f1} does not refer to any specific microscopic theory of black hole, general descriptions will be appropriate. In the following we show that the equality between the entropy of a black hole and its Bekenstein-Hawking entropy, 
\begin{equation}\label{f5}
S_{\text{black hole}}=S_{\text{BH}}=\frac{A}{4},
\end{equation}
can be obtained from the decoherence model.
Two remarks supporting \eqref{f5}  can be given: 

Firstly, with the relative radiation rate \eqref{555}, the total radiation rate is then
\begin{equation}\label{f7}
\Gamma=\prod_k\Gamma_k=\exp[-\int dM8\pi M]\triangleq e^{-\Delta S_{\text{BH}}},
\end{equation}
where $\Delta S_{\text{BH}}$ is the change in the Bekenstein-Hawking entropy. This is the same as that derived in the conventional approach \cite{Massar2000333}. 

Secondly, it has been shown in \cite{0264-9381-21-17-L02} that the Hawking temperature of a Schwarzschild black hole is just the minimal noise temperature $T_n$ added through the measurement of the black hole quasi-normal modes, 
\begin{equation}
T_n\geqslant T_{\text{BH}}.
\end{equation}
In the current case, this can be translated via \eqref{f4} (cf. also \cite{AH16}) into the information-theoretic context as
\begin{equation}
S_n\geqslant S_{\text{BH}}
\end{equation}
which, by  Schumacher's noiseless channel coding theorem for quantum information \cite{MikeIke}, means that the von Neumann entropy $S_{\text{BH}}$ of Hawking radiation measures the minimal physical resources to restore the information encoded in Hawking radiation. Indeed, the decoherence performs a projection from the original state to the diagonal density matrix, which can be envisioned as a data compression process provided that the whole system including the state and the observer is closed. Unitarity or information conservation of the total closed system requires the data compression to be reliable, for which the von Neumann entropy gives the minimal physical resources needed for the observer to decompress the data with reliable fidelity. (See the appendix for a construction of a possible data compression scheme for black holes.)
On the other hand, the noise temperature of a measurement is defined as the increase in the input temperature to account for the output noise \cite{0264-9381-21-17-L02}, while in the measurements on the outcoming squeezed states performed by an exterior observer, the noise temperature is effectively added from the black hole and the effect of the output noise turns out to be the entropy production due to the decoherence of the squeezed states.\footnote{If the black hole is taken as a quantum channel, then for the transmission of classical information in this quantum channel, the Holevo bound is positive \cite{AS14} and hence the classical information sent into the black hole could be recovered in principle. Now the Hawking radiations become noises to the channel. Since the channel capacity is positive, the noises cannot be bigger than the total information, so the best situation is where the noises are minimal. In fact, in \cite{BA14} it has been shown that the black hole quantum channel defined by the isometry \eqref{iso} and the output $\rho_{red}$ is a symmetric quantum channel with zero quantum capacity. Therefore, the Hawking radiations cannot carry quantum information and should be considered as noises.}
 This observation can be sketched as
\[\text{black hole}\xrightarrow{\text{entropy flow}}\text{Hawking temperature}\xrightarrow{\text{decoherence}}\text{entropy production}.\]
Hence, after obtaining the Hawking temperature from \eqref{f4}, one can apply the above arguments so as to translate the entropy of Hawking radiation to the entropy of black holes, since this noise temperature bounded from below by $T_{\text{BH}}$ represents the entropy flow from the black hole to the observer. An explicit quantum resources of information has been constructed as entanglement through a horizon in \cite{PhysRevD.34.373} where the entropy of a black hole has contributions from the exterior quantum fields (including the Hawking radiation) that takes the  area form \eqref{f5}, while the statistical-mechanical contributions depending on the number of internal quantum states are compensated and do not contribute to the thermodynamic entropy \cite{PhysRevLett.74.3319}. This justifies the above translation. Therefore, obtaining \eqref{f5} solely from decoherence is reasonable. Note that this way we can only assert that the black hole entropy is proportial to the horizon area $A$ with the coefficient, usally anticipated to be $1/4$, undetermined. This is in fact acceptible because for some black hole analogues in laboratory, such as acoustic black holes in BEC \cite{MR13}, their entanglement entropies for the pertinent reduced subsystem outside the horizon retain the area law but with coefficients different from $1/4$.

Next consider the state $\rho$ formed by \eqref{f33} interacting with a black hole. In analogy to \cite{1308.1245}, from $\rho$'s perspective its evolution can be described by a master equation in an open system,
\begin{equation}\label{master}
\dot{\rho}=i[\rho,H]+\mathbb{L}(\rho),
\end{equation}
where $\mathbb{L}$ represents the interaction with a black hole, the solution to which is a dynamical map or a superoperator $\$:\rho\rightarrow\$(\rho)$. For states in equilibrium with a black hole, one must have the stationary condition $\mathbb{L}(\$_e (\rho))=0$, while for the process of decoherence $\$_d$ must project onto the diagonal form
\begin{equation}\label{f10}
\$_d(\rho)=\sum_ip_i\ket{i}\bra{i}.
\end{equation}
In this model, one can discuss the thermodynamic properties by considering deviations from equilibrium, or equivalently the changes in the quantum coherence which can be measured by the relative entropy between states having coherence and the decohered states, as will be shown in next section.
\section{Four laws of black thermodynamics from decoherence}\label{3}
In this section, thermodynamic laws of black holes are derived from the thermodynamics of quantum coherence. 

The zeroth law of black hole thermodynamics \cite{BF01645742} states that the surface gravity $\kappa$ on the event horizon of a stationary black hole is constant so that $\kappa$ plays the role of  temperature $T$ in the usual sense. In order to obtain this from decoherence, one first notice that in \cite{1308.1245} the zeroth law of thermodynamics of quantum coherence is expressed as the vanishing of the relative entropy with respect to the equilibrium state,
\begin{equation}\label{f11}
S(\rho||\$_e (\rho))=\text{tr}(\rho\ln\rho)-\text{tr}(\rho\ln\$_e (\rho)).
\end{equation}
Similarly, for the decohered states $\rho_{red}$, by Klein's inequality \cite{MikeIke} we have
\begin{equation}
S(\rho||\$_d (\rho))\geqslant 0
\end{equation}
with equality if and only if $\rho=\rho_{red}=\$_d (\rho)$. Here the decoherence ignores the off-diagonal part of the original density matrix and thus ignores part of the original quantum coherence. Then by the monotonicity of relative entropy, $S(\rho||\$_d (\rho))$ decreases until the decoherence process ends at the decohered state $\$_d (\rho)$ where the quantum coherence has decreased to zero but the two relative states become identical. In other words, 
\[S(\rho||\$_d (\rho))\rightarrow S(\$_d (\rho)||\$_d (\rho))=0\]
just as the relaxation process in classical thermodynamics where the thermometer relaxes to equilibrium. In this respect, quantum states with coherence play the role of thermometers when coupled to a bath and then decohere to equilibrium, which in fact is more advantageous over the usual ones relying on relaxations \cite{1408.6967}. Hence for decohered states the Hawking temperature can be read off from \eqref{f4}, and in particular, for a Schwarzschild black hole, $T_{\text{BH}}=(8\pi M)^{-1}\triangleq\kappa/2\pi$, which recovers the original statement.

The first law of thermodynamics for a Kerr-Newman black hole is the geometric relation, $\delta M=(\kappa/8\pi)\delta A+\Omega\delta J+\Phi\delta Q$ \cite{BF01645742}. For a Schwarzshcild black hole, it is simply $\delta E_{\text{BH}}=T_{\text{BH}}\delta S_{\text{BH}}$ with $E_{\text{BH}}=M$. In the current situation, this geometric relation can be derived from the relative entropy. Indeed, it has been shown in \cite{10.1007/JHEP08(2013)060} that, since the relative entropy is a smooth nondegenrate function of states, if $S(\rho(\lambda)||\$_e (\rho))$ depends on the perturbation parameter $\lambda$ and $\rho(\lambda)=\rho(0)+\lambda\rho^\prime,~\$_e (\rho)=\rho(0)$, the first derivative of  $S(\rho(\lambda)||\$_e (\rho))$ vanish at $\lambda=0$, which entails
\begin{equation}\label{ff13}
\Delta S=\Delta\braket{H}
\end{equation}
for the first order variation of the entanglememt entropy $S$ and the expectation of the modular Hamiltonian ${H}=-\log(\rho)$. Now suppose two nearby equilibrium states are in the Hartle-Hawking states, $\$_d (\rho)\sim\exp[-E/T_{\text{BH}}]$\footnote{Under certain conditions the Hartle-Hawking state is a maximum of entanglemnet entropy \cite{PhysRevD.61.064015} and can be interpreted as the maximal amount of information transmitted from a black hole \cite{PhysRevD.58.104023}, which is desirable for the data compression arguments in section \ref{2}.}, with different energies. Then \eqref{ff13} gives
\begin{equation}\label{ff14}
\Delta S=\Delta\braket{H}=\frac{\Delta E_{\text{BH}}}{T_{\text{BH}}}\equiv\Delta S_{\text{BH}},
\end{equation}
which is just the first law of black hole thermodynamics for a Schwarzshcild black hole. For charged and spinning black holes, one can use the general Hartle-Hawking states with charges $Q$ and angular momenta $J$, $\$_d (\rho)\sim\exp[-(E-\Omega J-\Phi Q)/T_{\text{BH}}]$ \cite{HH76}. For states different from the Hartle -Hawking state, we can use an effective thermal model \cite{ELV10}: with the modular Hamiltonian ${H}=-\log(\rho)$, one can first write the density matrix of the squeezed state outside the black hole at equilibrium in the canonical form
\begin{equation}
\rho^{\text{eq}}_s=\frac{\exp[-\beta_s H_s]}{Z_s},\quad Z_s=\text{tr}\bigl(\exp[-\beta_s H_s]\bigr).
\end{equation}
Together with the black hole state and the correlation $C$, this gives a total density matrix $\rho^{\text{eq}}=\rho_s\otimes\rho_b+C$, with a total Hamiltonian applicable to von Neumann equations (instead of \eqref{master}),
\begin{equation}
H(t)=H_s(t)+H_b(t)+V(t),\quad V(t):\text{interactions}.
\end{equation}
Then the change in the entropy of the outside squeezed state can be written as
\begin{equation}
\Delta S=S(t)-S(0)=S(\rho_s(0)||\$_e (\rho_s))-S(\rho_s(t)||\$_e (\rho_s))+\Delta_eS(t)
\end{equation}
where the relative entropy terms equal the irreversible entropy (or correlation) production and the reversible entropy flow term $\Delta_eS(t)=\beta_s\Delta Q_s(t)$ is due to the reversible "heat" flow 
\begin{equation}
\Delta Q_s(t)=\text{tr}[\rho_s(t)-\rho_s(0)]H_s\equiv\braket{H_s}_t-\braket{H_s}_0
\end{equation}
from the black hole. Using that
\begin{align}
\text{tr}[H(t)\dot{\rho}(t)]=&\text{tr}\bigl[\bigl(H_s(t)+H_b(t)+V(t)\bigr)\dot{\rho}(t)\bigr]=\nonumber\\
=&\sum_i\bra{i}H(t)\rho H(t)-H(t)H(t)\rho\ket{i}=\sum_iE_i(t)\bra{i}H(t)\rho-H(t)\rho\ket{i}=0,
\end{align}
we have 
\begin{align}
\Delta Q_s(t)=&\int_0^tdt\frac{d}{dt}\bigl[\text{tr}H_s(t)\rho(t)\bigr]=\int_0^tdt\bigl[\text{tr}H_s(t)\dot\rho(t)\bigr]+\int_0^tdt\bigl[\text{tr}\dot{H}_s(t)\rho(t)\bigr]=\nonumber\\
=&-\int_0^tdt\bigl[\text{tr}(H_b(t)+V(t))\dot\rho(t)\bigr]+\int_0^tdt\bigl[\text{tr}\dot{H}_s(t)\rho(t)\bigr].
\end{align}
The work done on the total system is then
\begin{equation}
W=\braket{H(t)}_t-\braket{H(0)}_0=\int_0^tdt\text{tr}\bigl[\bigl(\dot{H}_s(t)+\dot{H}_b(t)+\dot{V}(t)\bigr){\rho}(t)\bigr].
\end{equation}
But in this isolated closed total system, the possible works are only done by the black hole charges and angular momenta with the effect on $\rho_s$ being $\Delta W=-W=\Omega\Delta J+\Phi\Delta Q$. Hence we see that the change, or more precisely the decrease in the energy of the black hole including the correlations is
\begin{equation}\label{first}
\Delta E=-(\braket{H_b(t)}_t-\braket{H_b(0)}_0)=\Delta Q+\Delta W=\beta^{-1}\Delta S+\Delta W,
\end{equation}
which represents the first law of black hole thermodynamics. From the term $\beta^{-1}\Delta S$ in \eqref{first} we see exactly how the temperature and entropy of the squeezed state outside the black hole are related to the black hole thermodynamics. Besides, the enenrgy in \eqref{first} includes the contribution from the correlations since the correlations usually resides at the black hole horizon. This accounts for the correction terms to the area term in the entropy formula.

The second law of black hole thermodynamics in the classical sense is the area law \cite{BF01645742}: $\delta A\geqslant0$, while after taking into consideration of Hawking effect it becomes the generalized second law \cite{PhysRevD.13.191}: $\delta(S_{\text{black~hole}}+S_{\text{m+r}})\geqslant0$. We want to derive these simply from from the knowledge of the squeezed states outside the black hole. At the first sight, the modular Hamiltonian that we have used generates a thermal time flow in the sense of Rovelli \cite{Rov93}, which explains why the of entropy of gravitational systems of this nonequilibrium kind grows in one of the directions of time.
In order to show the explicit inequalities, we recall that the second law of thermodynamics of any quantum process can be obtained by taking time derivitive of the relative entropy \eqref{f11} to get
\begin{equation}\label{f13}
\frac{d}{dt}S(\rho||\$_e (\rho))=\text{tr}(\dot\rho\ln\rho)-\text{tr}(\dot\rho\ln\$_e (\rho)),
\end{equation}
which is just the balance equation for entropies of nonequilibrium thermodynamics. Here $\sigma=-\frac{d}{dt}S(\rho||\$_e (\rho))$ is the rate of entropy production with the property that $\sigma=0$ for reversible processes and $\sigma\geqslant0$ in general, which is the local form of the second law of thermodynamics \cite{10.1063/1.523789}. For any quantum process, $\sigma\geqslant0$ is ensured again by the monotonicity of relative entropy that $S(\$(\rho)||\$_e (\rho))\leqslant S(\rho||\$_e (\rho))$ resulting in
\begin{equation}
\sigma=\lim_{t\rightarrow+0}\frac{S(\rho||\$_e (\rho))-S(\$(\rho)||\$_e (\rho))}{t}\geqslant0.
\end{equation}
The generalized second law also can be readily obtained from this monotonicity property \cite{0264-9381-25-20-205021}. Indeed, by assuming the decohered state to be again the Hartle-Hawking state and adopting the semiclassical quasistatic conditions in \cite{gr-qc/9705006v2}, one can see that the monotonicity of relative entropy gives
\begin{equation}
S(\rho||\$_d (\rho))-S(\$(\rho)||\$_d (\rho))\geqslant 0,
\end{equation}
which in this case becomes
\begin{equation}
S(\$(\rho))-S(\rho)-\frac{\braket{E}_{\$}-\braket{E}}{T}\geqslant 0.
\end{equation}
Using the first law \eqref{ff14}, one recovers (cf. also \cite{SW08})
\begin{equation}
\delta(S_{\text{BH}}+S)\geqslant0.
\end{equation}
Again, the Hartle-Hawking state can include other work terms, and for other states different form the Hartle-Hawking state one can use the effective thermal model as before to replace $\braket{E}/T$ by $\beta\braket{H}$.
To recover the classical area law, one notices that in the classical case where nothing could escape from a black hole the entropy flux term $-\text{tr}(\dot\rho\ln\$_e (\rho))$ in \eqref{f13} vanishes leaving 
\begin{equation}
\dot S=\sigma\geqslant0,
\end{equation}
which recovers the classical area law since we have obtained $S\propto A$ in the current decoherence model.

The third law of black hole thermodynamics \cite{PhysRevLett.57.397} states that the surface gravity $\kappa$ cannot be reduced to zero by any finite sequence of operations. This is of Nernst's form of unattainability principle, while in the current formulation one can provide a stronger Planck's form (or Nernst's theorem). Indeed, as $T_{\text{BH}}\rightarrow0$, the entropy of Hawking radiation \eqref{f4} vanishes, $S\rightarrow0$, which means that the in-vacuum state did not interact with the black hole and hence there is no entropy production in the outside states. With the noiseless channel coding theorem and arguments in the last section, the vanishing of $T_{\text{BH}}$ means the minimal resouces needed to store information are zero, that is, there is actually nothing to be stored. Thus no information inside the black hole has been transferred through Hawking radiation and the entropy of the black hole obtained from the knowledge of outside states keeps constant throughout, which can be taken as zero since the statistical-mechanical entropy depending on the number of internal quantum states do not contribute to its thermodynamic entropy \cite{PhysRevLett.74.3319}. This can be taken as the Planck's form of third law of black hole thermodynamics. Note that the extremal black holes obviously violate Nernst's theorem, but the above formulation only concerns with the observed thermodynamic entropy obtained via relation \eqref{f4} and hence respects Nernst's theorem. This is an effective way to extract correct Hawking temperature regardless of the corrections terms depending on other possible parameters such as charges and angular momenta \cite{Lin00}. 
\section{Discussion}\label{4}
The above derivations rely on the quantum coherence which is necessary for quantum entanglement. In fact, the squeezed vacuum state \eqref{f1} is an entangled state since it can be explicitly written as
\begin{equation}\label{3131}
\ket{ss_k}=\sum_{n=0}^{\infty}\frac{1}{\cosh r_k}(-e^{2i\phi_k}\tanh r_k)^n\ket{n_-}\otimes\ket{n_+},
\end{equation}
which is just the Schmidt decomposition of the Kruskal state into the Schwarzschild states separated by the horizon \cite{0264-9381-26-23-235008}. One might argue that the subsequent decoherence will destroy the quantum coherence and hence the entanglement. However, this does not destroy all the coherence or entanglement, and effectively describes the thermal nature in the Schwarzschild states observed by an exterior observer. Indeed, the entanglement spectrum of the Schmidt decomposition \eqref{3131},
\begin{equation}
p_n=\frac{\tanh^{2n}r_k}{\cosh^2r_k}
\end{equation}
is, just as the decoherence arguments given above, independent of the squeeze angle. The von Neumann entropy obtained from the limit of the R\'enyi entanglement entropy
\begin{equation}
S_1=\lim_{\mu\rightarrow1}\frac{1}{1-\mu}\ln\sum_{n}p^\mu_n=\lim_{\mu\rightarrow1}\frac{\ln(1-\tanh^{2\mu}r_k)+2\mu\ln\cosh r_k}{\mu-1}
\end{equation}
coincides with the von Neumann entropy \eqref{f44} for the decohered states. Then by identifying $p_n\sim \exp[\beta E_n]$, one can study the effective thermodynamics of the state \eqref{3131} \cite{SRP15}.

Note that in the above derivations in terms of squeeze states, unlike in traditional gravity researches, the usage of spacetime geometry has been reduced to a minimum. In this respect Hawking effect together with the black hole thermodynamics can be attributed solely to quantum decoherence. This underlies the laboratory analogues of black holes and Hawking radiation since both in black hole spacetimes \cite{BF02345020} and in laboratory systems \cite{0953-4075-45-16-163001} the derivation of Hawking effect only refers to the behaviors of wave propagations in a media with horizon. Therefore, we can say that for the black hole {\it analogues} in laboratory systems and for the gravitational black holes their Hawking effects are really the same thing in the sense that they have the same quantum description, say quantum decoherence. This also leads us to a  way of detecting Hawking radiations or black hole thermodynamics of the possible micro black holes produced in high energy colliders. A first indication of the decoherence effects in high energy processes has been recently pointed out in B meson decays \cite{Alok15}, so it would be interesting to investigate the gravitational decoherence effects pertaining to those micro black holes.

Of course, the approach of this paper is by no means an ultimate theory. But following this line of thought, we can actually find similar structures in quantum gravity models. For instance, acting a twist deformation on a D1D5 CFT state  results in a squeezed state deformed away from the orbifold point \cite{ACM10a}. Since in string theory the D1D5 CFT is the holographic dual of a near-extremal black hole, this deformation makes the original free theory at the orbifold point become an interacting theory with exitations that might be able to describe holographically the dynamical processes of black hole formation and evaporation \cite{ACM10b}. 
In this spirit, one can also take the gravitation as an emergent phenomenon from quantum physics, which, as a new implementation of Wheeler's {\it it from bit} proposal, gives a possible way to recover classical gravity from quantum gravity models. At this point, one can further reexamine many celebrated results in gravitational black-hole physics to find more quantum and informational nature of black holes. 

~

{\bf Acknowledgement}. The author thanks Prof. Ashutosh Kumar Alok for helpful suggestions.

\appendix
\section{More on black-hole data compression}
In this appendix, further discussions on the relation between data compression and Hawking radiation are given. Intuitively, one might expect the off-diagonal elements in the density matrix are compressed into the diagonal ones, just as the Schur-Weyl duality between computational basis and symmetric states. However, a concrete compression scheme depends on the  knowledge of specific data packing in a black hole.

In the conventional sense, information is expected to be distributed on the black hole horizon. But a different data packing scheme can be explored, as is done in \cite{doi:10.1142/S0218271814500412} recently. In \cite{doi:10.1142/S0218271814500412} data are packed in the light sheets of the entire interior holographic shells of a black hole, whose count procedure is analogous to the stack data structure in classical computer science. Using the binary number representing the Catalan parentheses construction (cf. Fig.3 of \cite{doi:10.1142/S0218271814500412}), one can envision the information release from black hole as discarding the outest shell, e.g.
\[
\ket{11010100}\rightarrow\ket{101010}\triangleq\ket{0}\ket{101010}\ket{0},
\]
where the original positions of the discarded shell (or the parentheses) are kept in the last notation.

With this notation, one can try to construct compression scheme in analogy to \cite{PhysRevA.81.032317} as follows. The computational basis is the basis consisting of the single shell $\{\ket{...1...}\}$, while the symmetic states are now those states containg various shells $\{\ket{...1...1...}\}$. Then a Schur-Weyl transform can be constrcuted in order to transform the symmetric states to the computational basis. The scheme in \cite{PhysRevA.81.032317} further compresses the $N$-qubit computational basis into the first $\log_2(N+1)$ qubits. Thus, one can envision that the diagonal matrix elements as the computational basis and the off-diagonal elements as the symmetic states, since the off diagonal elements contains the squeeze angles that can be related to the relative angles of different light sheets. In this respect Hawking effect as a decoherence effect of the squeezed states compresses the information into the (first $\log_2(N+1)$) diagonal elements of the density matrix, as is anticipated in section \ref{2}.

\bibliographystyle{elsarticle-num} 
\bibliography{guo}

\end{document}